# Technology, Propaganda, and the Limits of Human Intellect

Panagiotis Takis Metaxas
Computer Science Department
Wellesley College, USA
pmetaxas@wellesley.edu

ABSTRACT: "Fake news" is a recent phenomenon, but misinformation and propaganda are not. Our new communication technologies make it easy for us to be exposed to high volumes of true, false, irrelevant, and unprovable information. Future AI is expected to amplify the problem even more. At the same time, our brains are reaching their limits in handling information. How should we respond to propaganda? Technology can help, but relying on it alone will not suffice in the long term. We also need ethical policies, laws, regulations, and trusted authorities, including fact-checkers. However, we will not solve the problem without the active engagement of the educated citizen. Epistemological education, recognition of self biases and protection of our channels of communication and trusted networks are all needed to overcome the problem and continue our progress as democratic societies.

-------------------------------------

## Part 1. A major problem

### Introduction

Rarely a day passes that we do not hear about "fake news," but the term does not always mean the same thing to each of us. In this article, I use the phrase "fake news" (and stylize it in quotes) to refer to falsehoods online that are formatted and circulated such that a reader might mistake them for legitimate news articles [1].[1]

"Fake news" is a relatively recent phenomenon born out the financial attractiveness of "clickbait"—capturing users' attention online to generate revenue through advertisements. It has expanded into new forms of propaganda, bullshit (as defined by Harry Frankfurt [2]) and financial scams, all of which have existed since the creation of early human communities. Since ancient times people have been exposed to lies, misinformation, and falsehoods. Believing lies can come at grave personal and social costs. In some extreme cases, communities, religions, and cultures were destroyed due to misplaced belief in falsehoods. Even the end of the first Democracy came when the Athenian citizens, led by the charismatic populist leader Alcibiades, believed reports that their Syracusan enemies were cowards and Sicilian city-states would support them as allies, and entered in a disastrous war they could not win [3].

---
[1] As others have pointed out, some politicians, populists, and dictators misuse it to describe opinions they dislike; this co-opting of terminology is commonly done, as with other terms such as "patriotism" and "the people" [4].

As individuals, we have also been exposed to lies, misinformation, and false beliefs since the early stages of our lives. By adulthood, we corrected most (but likely not all) of them. We do not believe in Santa Claus anymore, but many of us still believe that the alignment of stars at one's time of birth has a significant effect on one's life. Some athletes and gamblers perform ceremonial procedures or keep a lucky object to which they attribute magical performance-enhancing powers.

Believing in lies may have negative effects at some point in our lives, though it is not predictable if and when such effects will happen for any particular lie. Today, we are exposed to "fake news," a newer phenomenon in the ever-widening genre of deception. Due to our communication technologies, the volume of misinformation reaching us comes at speeds and levels that we, as a species, have not evolved to handle.

## Personal perspective

I first became curious about the question of identifying truth as an undergraduate majoring in Mathematics at the University of Athens, Greece. A trusted friend, fluent in Russian, informed me that Soviet Mathematicians had devised a model by which one could determine the facts in news announcements. I tried to find more information about the model, but during the Cold War there was not much scientific communication between East and West, there was no Internet, and I could not read Russian. It took me years to realize that I had encountered an early instance of "fake news."

In 1994, when the first search engine appeared, I became interested in the phenomenon of online propaganda. As an educator, my concern developed when I realized that my students were using search engines to retrieve and use information without an understanding of how it was produced. This became apparent with the rise of Google Search, a technology that successfully presented itself being able to objectively measure information quality using its famous PageRank algorithm. In "Of course it's true, I saw it on the Internet," (2003) [5] we found that the more technically competent students were more likely to be fooled by unreliable search results because they trusted their ability to find information using this technological tool and applied less critical thinking. Until the advent of search engines, print information generally had greater validity over information gathered in other ways, so conducting quality research involved discovering printed sources—something that was quite hard, generally requiring hours spent pouring over library books. For something to get be published meant that it had been successfully scrutinized by professional editors, and challenged in ways that other forms of communication were not. But with the advent of the World Wide Web, in a sudden twist equal parts liberatory and dangerous, achieving publication became trivially easy. Barriers to authorship crumbled; in the Internet Era virtually anyone can be an author of content, and with the help of search engines, virtually any content authored can be found.

People trusted Google. Its PageRank algorithm was listed among the ten best algorithms of data mining [6], and its reputation was transferring to its search results. If you find an article in the top-10 search results, the thinking goes, it must be good. Sure, you could encounter a bad search result, but you would think that it was due to the lack of your searching skills and you would try again with different keywords. But when you were searching for things you had no prior knowledge, there was no way of recognizing misinformation. And, as an author, if your article is not easily found with a search engine, it may as well not exist. Most people do not know (even today) that search engines can be gamed to promote the page of an experienced manipulator, or Web spammer. "Search Engine Optimization" a 65 billion dollar industry born in the early 2000's was making a living by fooling Google and the other search engines on particular search queries. Propagandists and advertisers jumped at the opportunity. In "Web Spam, Propaganda and Trust" (2005) [7] we describe some of the ways that manipulation ("Web spam") is done.

And, while Google has been successful in defending itself against attacks against electoral candidates since 2008, Twitter and Facebook have become the new battleground. We observed the first Twitter bomb during the 2010 MA special senatorial election [8], and we observed that the same technique was employed on Facebook during the 2016 US elections [9]. Propagandists create "fake news" and fake accounts, then infiltrate and lurk for a while at the sidelines of political groups, especially those who feel angry and emotionally charged forming echo chambers. At some appropriate time, the fake accounts start promoting their "fake news" to the group and watch them spread in their echo chambers. After that, the fake accounts can delete themselves, making it difficult even to determine who was the initiating propagandist. The infamous "Pizzagate" conspiracy theory is an example of such behavior that we observed and documented [10].

Search engines and Social media platforms have succeeded in gaining the trust of their users. Willing or not, they are the de facto arbiters of what information users see [7]. And they are under continuous attacks to promote the agenda of a propagandist, an advertiser, a fanatic or a conspiracy theorist.

With "fake news" propagating at the time of elections, however, the stakes have been extremely high. Messing up with elections means manipulating one of the pillars of Democracy. Avoiding this manipulation has to be a coordinated effort. In particular, both Social Media platforms and citizens need to do a better job of recognizing "fake news" online. Since the 2016 Brexit and US elections, most attention has fallen on the platforms, and rightly so.  What can we, the citizens, do?

Our first collective reaction was to demand the deployment of fact-checking procedures. We want fact-checkers that will reliably determine the truth behind any online rumor and will notify us immediately. We want our social networks to inform us of what is reliable and what not. We want policies that will enforce trustworthy information and laws that will punish those who violate them. And we want magical technologies to apply immediately, automatically, correctly every time.

In other words, we demand benevolent censorship.

Unfortunately, this cannot happen. There is no universal agreement on what is true and what is not. One person's spam can be another person's treasure.  Laws are likely to be outdated by the advance of technology by the time they pass. Policies will be a weapon against the public in the authoritarian regimes that we see appearing around the world these days, as the new Malaysian "Anti-Fake News Bill 2018" demonstrates [11]. Partisan fact-checking will not be trusted. Independent fact-checkers such as Academic Librarians, who enjoy broader support, are unwilling to step into the ring, and understandably so. Much promising deep learning technologies are as likely to help as to hurt, as the so-called "deep fake" videos already demonstrate [12]. Moreover, machine learning technologies that do not give evidence on how they make decisions can only learn and imitate biases and injustices, as we have already done in our communities [13].

This does not mean we should not deploy dependable, independent, experienced, educated smart crowds to help evaluate "fake news." We should. We should encourage Academic Librarians to take the role of overseeing fact-checking organizations. Social networks and search engines should diminish the financial incentives that drive the "Web spam" and "fake news" producers. User interfaces need to be clearer to help us detect "click bait." But, nothing will succeed, unless we, the consumers of information, take more responsibility. We cannot avoid it any longer.

## Part 2. Epistemological education

How easy is it for people to recognize "fake news"? How do we know what we know? This is one of the fundamental questions that we need to answer to evaluate the performance of any fact-checking system, operated by humans or machines. There is a whole branch of philosophy, called epistemology, that deals with the establishment of knowledge. Briefly, we know things due to

(a) our own experiences,
(b) our trust in reliable authorities, and
(c) our personal skill in handling logical reasoning.

The latter one is often called "critical thinking skills." And each of these three sources of knowledge (intrinsic, extrinsic and derivations) is challenged by our technologies today.

We hold some beliefs for extrinsic reasons: because we trust the entity that is providing or supporting the information. Our whole educational system is based on this premise. Our systems of governance are also based on this, especially the "Fourth Estate," the Press. ("Of course it's true, I saw it in the newspaper!" was an expression that people in my home country used to use when they wanted to support a claim they believed as true. It is this expression that inspired the title of the article I mentioned above [5]).

We also hold some beliefs for intrinsic reasons, that is, based on our own experiences realized through our senses and interpreted with our mind. We trust our senses. We consider our experiences fundamental, and we rarely question whether what we learn from them is ever in doubt. "I saw it with my own eyes" is an expression that exists in most, if not all, languages. Trusting our eyes is considered equivalent to having absolute confidence since seeing is the most powerful of all our senses.

But it rarely is the case that we can make sense of what we see or hear without some thought process that interprets and establishes the factual aspects of our experience. Here is where we need the support of critical thinking skills that will help us determine the validity of our thoughts and observations. And for that, we also want the aid of a sound mind.

Even though in daily conversations we use the term "critical thinking" as a synonym to "common sense," they are quite different, as Duncan Watts explains in "Everything is Obvious once you know the answer" [14]. Critical thinking means to use mathematical logic and rigor, to combine things that you already know, to derive new knowledge. By rigor, we mean to apply the Scientific Method in the derivation, and it starts with writing down carefully a hypothesis. Committing the hypothesis on a fixed medium is a crucial step because without it, our thoughts may drift and we may end up evaluating something quite different. Then, we need to search for evidence, both supporting the hypothesis *and* discrediting it. Looking for both types of evidence is essential because searching only for confirming evidence and ignoring discrediting evidence is the basis of most fallacies and conspiracy theories.

Of course, we need to apply mathematical logic in making sense of the evidence we have collected. This technical part of evaluating the hypothesis is hard, as it requires both education and practice. Unfortunately, not all educational systems prepare people for this step. Logic rules can be confused. A common logical mistake often used in conspiracy theories is to consider lack of evidence against a hypothesis as proof of correctness. If I cannot see why something could be wrong, the erroneous thinking goes, it must be true. For example, Facebook used flags to denote that some news article has been fact-checked as true [15]. But flagged articles are few, and most articles are not checked. Someone mistakenly may derive that an article without a flag is false. However, it is more likely that no one has checked that second article. (The "closed world" assumption does not hold for the fact-checked articles.)

Unfortunately, applying the Scientific Method all the time is not easy for three reasons.

The first reason relates to effort and education. It is mentally tiring, and it requires training to think critically. The good news here is that the more you practice critical thinking, the less tiring it becomes, and you can even come to enjoy it.

The second reason relates to prejudices: we need to be aware of our own biases. We need to have what ancient Greek philosophers termed "γνωθι σ'αυτον" or self-knowledge. We are not, by default, the objective judges we may wish to be. There are many cognitive biases we carry along, and they are all transparent to us. One of the more commonly used is confirmation bias: When we are presented with facts, it is easier to cherry pick those that agree with what we already believe and discredit those that do not, than changing our opinion [16]. The longer we think in a certain way, the more we reinforce the existing neural connections, the harder it is to change them.  Changing our thinking requires effort and time.

The third reason is related to false information we have already accumulated over the years. Not everything we were taught as children by parents, teachers and the community, is true. Some of the "facts" we learned were made up stories easy to comprehend, and provided comfort for our worries. Belief in Astrology is just one such example. Even though it is relatively easy for people to see it as invalid, many choose to believe in it and have it guide some of their actions. (I call it "relatively easy" compared to some religious beliefs, which are incredibly hard to see in our own religion–though they are easy to spot in other people's religious beliefs.)

We also know "facts" that are related to brain's limitations, for example, "facts" that we misunderstood or misheard. We may also remember incorrectly. The brain is not a reliable database of information that can be accessed accurately and on demand every time. Early efforts to understand the brain were comparing it to computer memory, but this metaphor is misleading. Every time we remember something, we are reconstructing the memory through an inexact and unreliable process. Sometimes police detectives and lawyers are successful in implanting "memories" to unsuspected witnesses by describing in detail the events. Witnesses may end up "remembering" these "memories."

We may also know "facts" because we applied the wrong pattern trying to make sense of something. Our brain is a pattern-matching machine. On the one hand, finding similarities and patterns is fundamental to creativity. We develop solutions by recognizing similarities between a situation we are familiar with, and another that is new to us. But, on the other hand, our pattern-matching ability can fail us sometimes: We can see the image of a face on a rock in NH or on the surface of the Moon the moment someone points it out.

We also know "facts" that we have observed under emotional stress, such as fear, anger, and passion, as we often feel during important political elections. (And, of course, we sometimes know "facts" that we derived upon when our brain was not working reliably. This could be due to the influence of alcohol or other chemical substances, due to lack of sleep, due to mental illnesses, or even due to extreme focus, as with the famous case of missing the gorilla appearing in a video of people passing a basketball.)

I am sure that the reader can add more such examples that they have observed in others. It is always far easier to observe such behavior in others because, well, if we realize that they are happening to ourselves, we could try to correct them, assuming we have enough mental power for that.

What's wrong with our brain?

Our brain is impressive, but it is not perfect, and it does not work well every time or at every phase of our life. It is a very complex organ. Our brains are the products of evolution. They are not perfect or complete; they are work-in-progress. Neuroscientists describe the evolutionary process that started with the so-called "reptilian brain" the smaller component that responds immediately to the basic instincts: fear, hunger, sexual desire [17]. (Kahneman's "System 1" may have its headquarters in that part of the brain [18].) On top of that we have the mammal brain, then the primate brain, then the human brain occupying much of the neocortex. (This last one may be primarily responsible for Kahneman's "System 2".)

Our brain is affected by construction limitations and errors. Our feelings, our senses, our environment challenge our perception of reality. We need to feel that we are in control of our environment to survive. The natural world around us is full of randomness, but we do not readily accept randomness in phenomena, we want to "discover" reasons explaining randomness. Again, this desire for control and for an explanation that discounts randomness is a powerful source for conspiracy theories.

Are we stupid?

No, we are not stupid; we are thinking lazy. Our brain has a hard time staying focused for too long [19]. When we sit down to study intensely for an hour or two, we end up feeling mentally and physically exhausted, though we did not move from that chair. Thinking critically is very taxing to the brain. We try to avoid it unless we have to, like when we take a test in school. In most other cases, we try to create shortcuts to avoid using all of it. We adopt heuristics, stereotypes, personal ways of "thinking" that most of the time serve us well – but not always [20].

It is counter-intuitive, but even binary logic does not always help. Since a young age, we practice with statements that are either true or false, because they are easier to understand. We are so accustomed to the process of understanding our world through a "closed world" assumption: We often assume that if something cannot be shown to be true, it must be false. Well, it may be impossible to determine its validity under the accepted assumptions. (This is a belief that Mathematicians had until the early 20th century when Gödel proved that axiomatic mathematical systems containing basic arithmetic are incomplete [21]. They contain many mathematical statements that one can neither prove they are correct nor prove incorrect. In our educational systems, however, we are never given exercises for which we cannot have a definite answer. )

Given our millennia-long practice with a binary system of logic, it is not surprising that human languages have evolved without expressions for ambiguity or probabilities. We find it normal to start sentences with "Women are ..." as if it is even possible to make a statement that applies to all women (and be upset when we see Google Suggest providing insulting completions). And we have no problem understanding what the sentence "This bar is so crowded, no one ever goes there" means. I would challenge the reader to find even one statement about human social behavior that is always true.

We have been dealing with this situation for hundreds of years. Our technologies have helped us make progress in controlling the world around us, but have also challenged us.

Consider one of the most impressive technologies of all times, the technology of writing. Using little drawings to depict phonemes, words has been one of the more profound technologies of all times. We spend a good time of our lives training to recognize words, form sentences, compose arguments. Writing has enabled us to transmit ideas and information across generations. Every time we made it more efficient, as with the invention of the printing press, it had a profound effect on human history. But up until the spread of the interconnected networks of social media, we had few books to read. Few, of course, compared to the tsunami of words we read these days on Facebook, online newspapers, and, well, fake newspapers. Nowadays, AI programs can write simple articles, such as game reports from sporting events. Already AI algorithms can do a very good job in translating automatically from any language to our own. The amount of information that reaches us has exploded. But our attention and time has remain the same.

Censorship used to be the act of hiding information from the public. In our new world, overwhelming noise can also function as censorship, as "epistemic paralysis" in Jonathan Zittrain's terms [22]. Our social media are shifting our attention constantly between issues and topics with such speed and volume that we lose track of what is important.

## Part 3. No Easy Solutions

The last couple years have seen a massive rise in interest in "fake news." But although many researchers, politicians, lawmakers, and laypeople are trying to address this issue, it remains an immensely complex problem, challenging the limits of our human intellect. What can we do?

Technology can certainly help, primarily if it is used to discover valid evidence that diverges from what we are casually aiming to locate. It can also help to inform us when we find ourselves in an echo-chamber. Wiki technology maintaining evidence considered by fact-checkers and librarians can also help. It can help them to manage the process and the rest of us to monitor what led to their decisions. Interfaces that give comparable exposure to a claim and its refutation would also help, so that, e.g., the comments refuting a claim appearing in a social medium are reasonably visible.

Laws, regulations and policies can also help: Laws limiting the financial incentives that enable pranksters, propagandists, and advertisers in producing misinformation to draw our attention and clicks. Regulations that restrict the collection and exchange of personal information, such as the recent European GDPR[2]. Ethical policies that protect our limited attention capital so that we do not sink in the constant wave of information that reaches us every day.

Policies, laws, regulations, trusted authorities, and technology will help, but they will not solve the problem of propaganda, misinformation, and "fake news" if we just rely on them. No solution will be complete without active engagement of the citizen, epistemological education (not just media literacy [23]), and an active democratic society. We need to be aware of why we believe what we believe and of our own biases. We need to listen to those outside our echo chambers. And we need to apply critical thinking habitually so that we enjoy practicing it on a daily basis. If we adopt a solution that requires even less use of our brain, we may just become irrelevant [24].

The above is a long list of challenging skills. Mastery of (most of) these skills is necessary for a successful life in the 21st century and beyond. It is not easy because we need to change ourselves. But that's what Education was always about.

# References

[1] "Separating Truth from Lies" (email interview), by Alison Head and Kirsten Hostetler, Project Information Literacy, Smart Talk Interview, no. 27 (21 February 2017)
http://www.projectinfolit.org/takis-metaxas-smart-talk.html

[2] Harry Frankfurt, "On Bullshit."
Princeton, NJ: Princeton University Press, 2005.

[3] Mary Lefkowitz, "Do Facts Matter?
Redefining truth is a tried and true method of taking control." The Spoke, Albright Institute for Global Affairs, Feb 24, 2017.
https://www.wellesley.edu/albright/about/blog/3261-do-facts-matter

[4] "The media's definition of fake news vs. Donald Trump's", Politifact,
By Angie Drobnic Holan on Wednesday, October 18th, 2017
http://www.politifact.com/truth-o-meter/article/2017/oct/18/deciding-whats-fake-medias-definition-fake-news-vs/

---

[2] Wikipedia entry of EU's General Data Protection Regulation http://bit.ly/2xfqqb3

[5] L. Graham and P. Metaxas, 'Of course it is true; I saw it on the Internet. Critical thinking in the Internet Era."
Communications of the ACM, May 2003.
http://bit.ly/oMjgnw

[6] X. Wu et. al.,
"Top 10 algorithms in data mining", Knowl Inf Syst (2008) 14:1–37
http://www.cs.uvm.edu/~icdm/algorithms/10Algorithms-08.pdf

[7] P. Metaxas and J. Destefano, "Web Spam, Propaganda and Trust", in Adversarial Information Retrieval (AIRWeb), WWW 2005 Conference, Chiba, Japan.
http://airweb.cse.lehigh.edu/2005/metaxas.pdf

[8] P. Metaxas and with E. Mustafaraj., "From Obscurity to Prominence in Minutes: Political Speech and Real-Time Search" Web Science 2010 Conference, Raleigh, NC, April 2010.
http://bit.ly/Twitter-Bomb

[9] E. Mustafaraj and P. Metaxas, "The Fake News Spreading Plague: Was it Preventable?" in WebScience 2017, Troy, NY, June 2017.
http://bit.ly/2sehUCv

[10] P. Metaxas and S. Finn, "The infamous "pizzagate" conspiracy theory: Insights from a TwitterTrails investigation"
Computation and Journalism 2017
http://bit.ly/2xEfIKU

[11] "Malaysia's anti-fake news law raises media censorship fears"
By Marc Lourdes, CNN, April 3, 2018
https://www.cnn.com/2018/03/30/asia/malaysia-anti-fake-news-bill-intl/index.html

[12] "Here come the Fake Videos, Too,", By Kevin Roose
NYTimes
March 4, 2018
https://www.nytimes.com/2018/03/04/technology/fake-videos-deepfakes.html

[13] "Bias in Criminal Risk Scores Is Mathematically Inevitable, Researchers Say"
by Julia Angwin and Jeff Larson
Dec. 30, 2016
https://www.propublica.org/article/bias-in-criminal-risk-scores-is-mathematically-inevitable-researchers-say

[14] "Duncan Watts, "Everything is Obvious (once you know the answer). How common sense fails us". Crown Business; First Edition edition (March 29, 2011)
http://everythingisobvious.com/the-book/

[15] "Facebook will stop labeling fake news because it backfired, made more users believe hoaxes"
By Jason Silverstein, 12/21/17
http://www.newsweek.com/facebook-label-fake-news-believe-hoaxes-756426

[16] "The Nature and Origins of Misperceptions: Understanding False and Unsupported Beliefs About Politics"
By D.J. Flynn, Brendan Nyhan, Jason Reifler
Advances in Political Psychology (38):S1
February 2017

[17] "Thinking fast, and slow", Daniel Kahneman.
Farrar, Straus and Giroux
2011

[18] "A machine for jumping to conclusions."
By Lea Winerman
Monitor on Psychology, APA,
February 2012, Vol 43, No. 2

[19] Does Thinking Really Hard Burn More Calories? By Ferris Jabr on July 18, 2012
Scientific American https://www.scientificamerican.com/article/thinking-hard-calories/
Alt[19] http://worldview.stanford.edu/blog/ask-neuroscientist-why-thinking-hard-so-hard

[20] Cathy O'Neil, "Weapons of Math Destruction: How Big Data Increases Inequality and Threatens Democracy"
Penguin Books, Limited, 2016

[21] Wolfgang Ertel, "Introduction to Artificial Intelligence"
Springer-Verlag London, 2011
https://books.google.com/books?id=geFHDwAAQBAJ&pg=PA68

[22] Jonathan Zittrain interview at the Albright Institute, January 2018,
https://www.youtube.com/watch?v=lrRjLiM0wTI

[23] danah boyd 2018 SXSW EDU Keynote "What Hath We Wrought?"
https://www.sxswedu.com/speaker/danah-boyd/

[24] Yuval Noah Harari, "Homo Deus: A Brief History of Tomorrow".
HarperCollins Publishers, 2017. http://www.ynharari.com/book/homo-deus/